\documentclass[aip,jcp,reprint]{revtex4-1}
\usepackage{epsfig}
\usepackage{color}
\usepackage{amsmath}
\usepackage{amsfonts}
\usepackage{natbib}
\usepackage{notes2bib}

\newcommand{\degree}{^{\circ} }

\newcommand{\avg}[1]{\left<#1\right>}

\newcommand{\brac}[1]{\left[#1\right]}

\newcommand{\ie}{\emph{i.e.}}

\newcommand{\etal}{\emph{et al.}}

\begin{document}


\title{Halogen Bond Structure and Dynamics from Molecular Simulations}


\author{Richard C. Remsing}
\email[]{rremsing@temple.edu}
\author{Michael L. Klein}
\email[]{mike.klein@temple.edu}
\affiliation{Institute for Computational Molecular Science and Department of Chemistry, Temple University, Philadelphia, PA 19122}




\begin{abstract}
Halogen bonding has emerged as an important noncovalent interaction in a myriad of applications,
including drug design, supramolecular assembly, and catalysis. 
Current understanding of the halogen bond is informed by electronic structure calculations on isolated molecules and/or crystal structures that are not readily transferable to liquids and disordered phases.
To address this issue, we present a first-principles simulation-based approach for quantifying halogen bonds
in molecular systems rooted in an understanding of nuclei-nuclei and electron-nuclei spatial correlations.
We then demonstrate how this approach can be used to quantify the structure and dynamics of halogen bonds
in condensed phases, using solid and liquid molecular chlorine as prototypical examples with
high concentrations of halogen bonds.
We close with a discussion of how the knowledge generated by our first-principles
approach may inform the development of classical empirical models, with a consistent representation of halogen bonding. 
\end{abstract}


\maketitle

\raggedbottom

\section{Introduction}

Noncovalent intermolecular interactions manifest the major driving forces
in a wide variety of physicochemical processes.
In this context, most attention has been focused on steric repulsion, 
hydrogen bonding, and van der Waals interactions.
But in recent years, halogen bonding has emerged as a prominent interaction
in many applications~\cite{Cavallo_2016,Resnati:AccChemRes2005,Resnati:AccChemRes,Desiraju:AccChemRes,Politzer_2010,Voth_2009}.
In particular, halogen bonds have been successfully utilized in
crystal engineering~\cite{Desiraju:AccChemRes,Tordin:CrystGrowthDes,Evans_2019},
self-assembly~\cite{Resnati:AccChemRes,2DXB-Science},
and to tune reactivity in synthetic and catalytic applications~\cite{Bulfield_2016,Metrangolo_2018}.
Biomolecular halogen bonds have also been exploited to enhance protein-ligand binding
strengths~\cite{Parisini_2011,Auffinger_2004,Ford_2016} and biomolecular assembly~\cite{Voth_2007},
and promise to play an important role in the future of therapeutics~\cite{Ford_2016,Carlsson_2018}.
Despite the importance and promise of halogen bonds in
chemistry, biology, and materials science, an atomic level quantification is lacking.
In this work, we use electronic structure calculations to develop a quantitative halogen
bonding definition in condensed phases, using both nuclear and electronic correlations
by exploiting maximally localized Wannier functions~\cite{RevModPhys.84.1419}.
We then employ \emph{ab initio} molecular dynamics (AIMD) simulations
and use this definition to analyze halogen bonding in solid and liquid Cl$_2$,
model systems with a high concentration of halogen bonds (XBs). 
We also discuss how our results can inform the description
of XBs within classical empirical models that are able to reach larger length
and time scales than the AIMD simulations.
%

\section{Simulation Details}
All calculations employed the \texttt{CP2K} package, 
and energies and forces were evaluated using the \emph{QUICKSTEP} module~\cite{VandeVondele2005}.
\emph{QUICKSTEP} employs basis sets of Gaussian-type orbitals and plane waves for the electron density,
leading to an efficient and accurate implementation of DFT~\cite{VandeVondele2007}.
We employ the molecularly optimized (MOLOPT) Godecker-Teter-Hutter (GTH) triple-$\zeta$,
single polarization (TZVP-MOLOPT-GTH) basis set~\cite{VandeVondele2007}
and the GTH-PADE (LDA-based) pseudopotential~\cite{Goedecker1996}
to represent the core electrons.
The valence electrons were treated explicitly, using the PBE~\cite{PBE} or BLYP~\cite{BLYP,BLYP2}
functionals as implemented in CP2K,
or the SCAN functional~\cite{SCAN,SCANNature} as implemented in LibXC~\cite{MARQUES20122272,LEHTOLA20181},
with a plane wave cutoff of 400~Ry.
The D3 van der Waals correction of Grimme~\etal \ was employed with the PBE and BLYP functionals,
as implemented in CP2K~\cite{Grimme},
and the rVV10 van der Waals corrections parameterized for use with SCAN
was employed to correct the SCAN functional~\cite{HaoweiPRX}.
Equilibration to a constant temperature was achieved by using a Nos\'{e}-Hoover
thermostat chain of length three~\cite{Nose_1984,Nose_1984b} with an integration timestep of 1.0~fs.
Systems were then further equilibrated in the microcanonical (NVE) ensemble for at least 10~ps,
before gathering statistics over at least 4~ps of production simulation time.
Maximally localized Wannier functions (MLWFs) were obtained using CP2K, minimizing the spreads
of the MLWFs according to the formulation of Ref.~\onlinecite{PhysRevB.61.10040}.
%

\section{Ab Initio Structure and Dynamics of Liquid Chlorine}

\begin{figure}[tb]
\begin{center}
\includegraphics[width=0.43\textwidth]{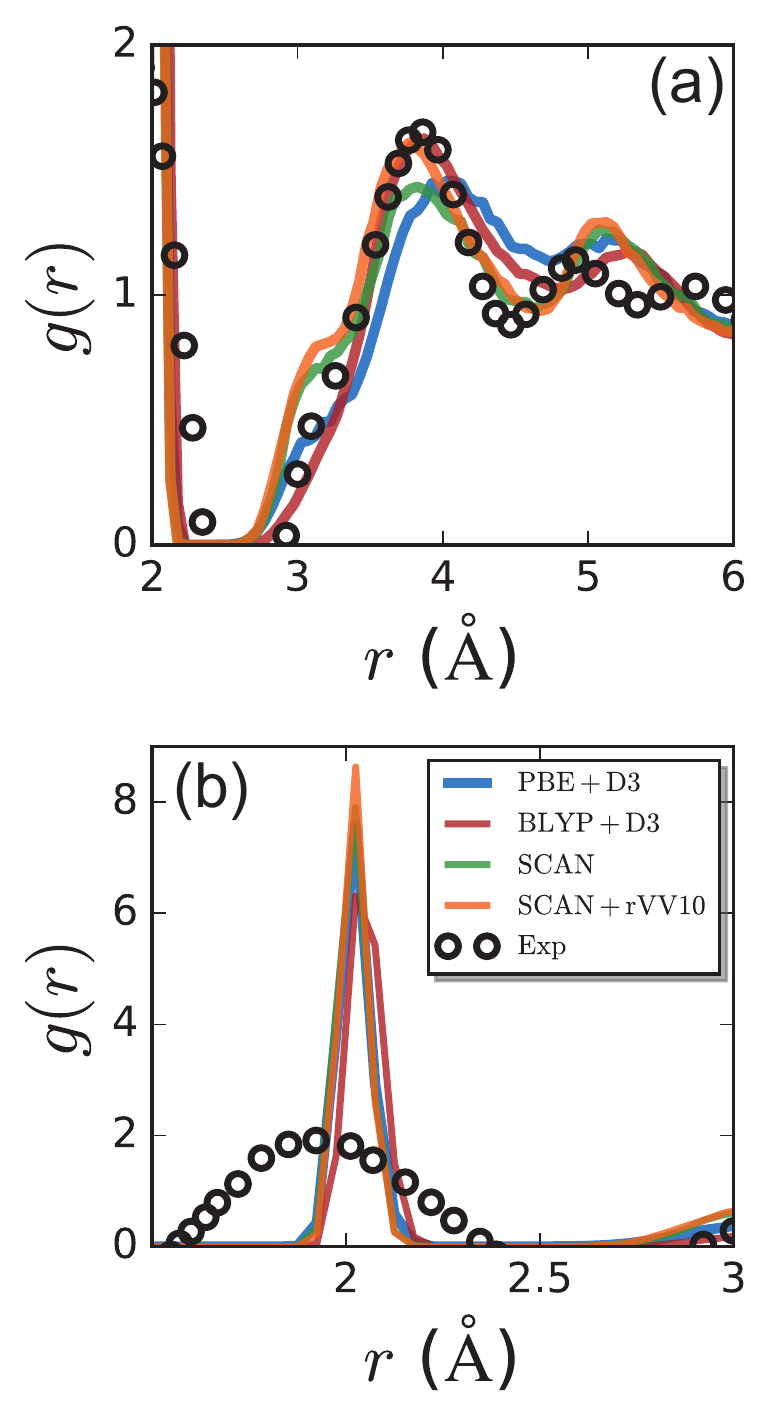}
\end{center}
\caption
{
Radial distribution functions, $g(r)$, for Cl-Cl correlations in liquid Cl$_2$ at $T=200$~K
and $\rho=12.5$~molecules/nm$^{3}$ as determined by neutron scattering measurements~\cite{Ricci:1983}
and four different density functional approximations.
Both (a) intermolecular and (b) intramolecular correlations are shown.
}
\label{fig:rdfcomp}
\end{figure}

Before quantifying halogen bonding in condensed phases of Cl$_2$, we evaluate
the ability of DFT-based approaches to predict the structure and dynamics of liquid chlorine ($l$-Cl$_2$).
We first focus on the structure of $l$-Cl$_2$ as quantified by the radial distribution function, $g(r)$,
shown in Fig.~\ref{fig:rdfcomp} for $T=200$~K and a density of $\rho=12.5$~molecules/nm$^3$,
as obtained from simulations and experimental neutron diffraction measurements~\cite{Ricci:1983}.
The position and height of the first major intermolecular peak in $g(r)$, as well as the first minimum,
are best captured by the SCAN+rVV10 description of Cl$_2$, with SCAN and BLYP+D3 also providing
a reasonable description of the liquid structure.
The PBE+D3 functional shifts the first intermolecular peak to larger distances.
All functionals yield a poor description of the second peak in $g(r)$.
This is due to an overestimation of the Cl-Cl bond length; all estimate this distance above 2~\AA,
in contrast to the experimental bond length of 1.99~\AA, see Fig.~\ref{fig:rdfcomp}b.
The simulated $g(r)$ displays a shoulder near 3~\AA, the amplitude of which is dependent on the functional.
As discussed in more detail in subsequent sections, this shoulder arises from halogen bonded dimers.
Thus, the height of this shoulder is proportional to the strength of XBs in each system.
The height of this shoulder, and consequently the XB strength, follows
SCAN+rVV10 $>$ SCAN $>$ PBE+D3 $>$ BLYP+D3.
As compared with the experimental $g(r)$, SCAN+rVV10 and SCAN overestimate this shoulder,
consistent with recent work showing that SCAN-based approaches can overestimate the strength of halogen bonds~\cite{Kim_2018}.
The PBE+D3 functional yields a reasonable description of the shoulder, despite
a worse description of subsequent intermolecular correlations.
BLYP+D3 underestimates the magnitude of this shoulder, and therefore the strength of halogen bonds
in this system.
%

\begin{figure}[tb]
\begin{center}
\includegraphics[width=0.43\textwidth]{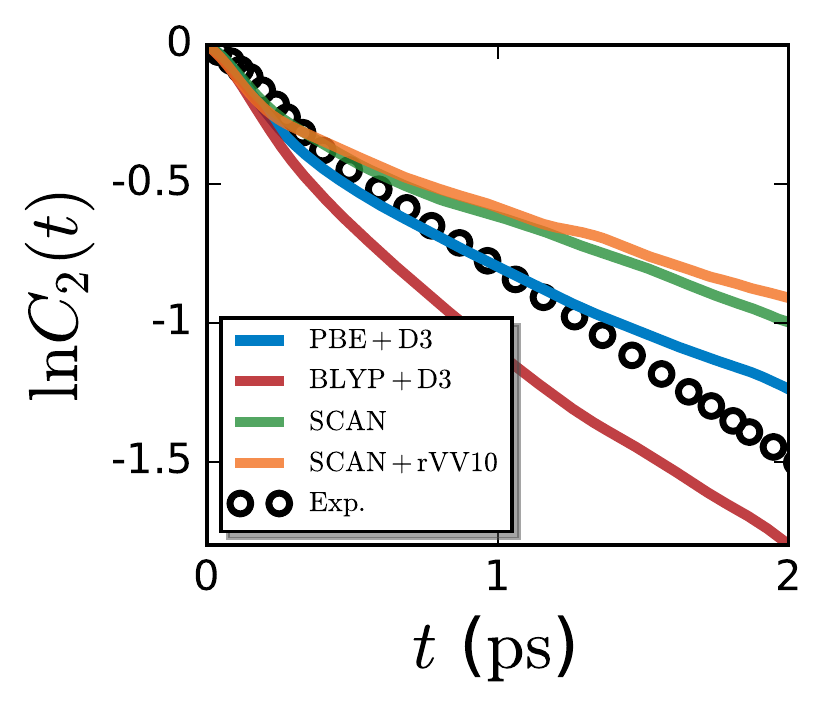}
\end{center}
\caption
{
Rotational correlation function, $C_2(t)$, as determined through Raman spectroscopy (Exp)~\cite{ClRaman}
and as predicted by the four density functional approximations used here.
Experimental results were obtained at 198~K.
}
\label{fig:rotcor}
\end{figure}

%
Dynamic properties also provide a stringent test of ab initio predictions.
In particular, we compare our predictions of the rotational time correlation function (TCF), $C_2(t)$, to experimental results,
where
\begin{equation}
C_2(t)={\avg{P_2(\mathbf{u}(t)\cdot\mathbf{u}(0))}},
\end{equation}
$P_2(x)$ is the second order Legendre polynomial
and $\mathbf{u}(t)$ is the Cl-Cl bond unit vector at time $t$.
The rotational TCF $C_2(t)$ and its associated rotational correlation time $\tau_2$ can be determined
experimentally through Raman~\cite{ClRaman} and NMR spectroscopy~\cite{Obermyer_1973}.
We compare our predictions to results from Raman spectroscopy~\cite{ClRaman} in Fig.~\ref{fig:rotcor}.
%

%
The rotational relaxation of $l$-Cl$_2$ is intimately tied to halogen bonding. 
In particular, if a chlorine dimer is halogen bonded with a neighbor, that XB must be broken
in order for the dimer to rotate by a significant amount (ignoring the possibility of the pair rotating collectively with the bond intact).
A similar rotational relaxation mechanism is known for water and other liquids with directional attractive interactions;
\ie \ H-bond breakage is involved in the rotational relaxation pathway of water~\cite{Laage_Sciene_2006,Laage_2008}.
Thus, we expect rotational relaxation to provide a sensitive, albeit indirect, probe of halogen bonding in $l$-Cl$_2$.

\begin{figure}[tb]
\begin{center}
\includegraphics[width=0.43\textwidth]{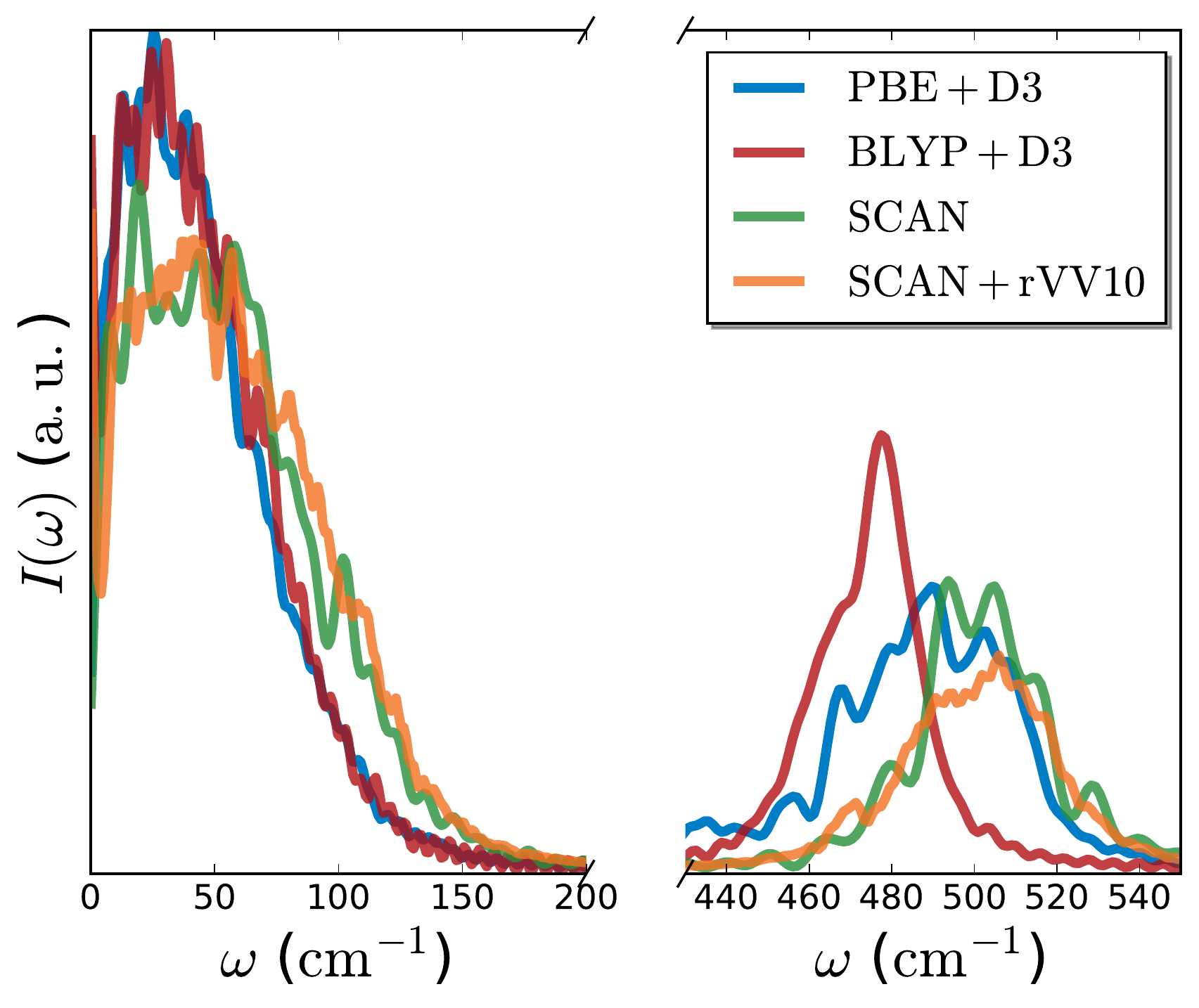}
\end{center}
\caption
{
Vibrational density of states, $I(\omega)$, as predicted by the four density functional approximations used here.
}
\label{fig:pwr}
\end{figure}

Indeed, the description of $C_2(t)$ provided by each functional closely tracks their respective ability
to capture the shoulder in $g(r)$ at close distances assigned to halogen bonded dimers.
The PBE+D3 functional yields an accurate description of the rotational dynamics
of $l$-Cl$_2$, as described by $C_2(t)$.
Both SCAN and SCAN+rVV10 yield rotational dynamics that are too slow, due the larger barrier to breaking XBs
in these systems.
The BLYP+D3 functional predicts a $C_2(t)$ that decays much too fast, consistent with the above conclusions that BLYP+D3
yields weaker XBs than expected.
Finally, we compute the vibrational density of states, $I(\omega)$, as the Fourier transform of the 
velocity autocorrelation $C_v(t)$, given by
\begin{equation}
C_v(t)=\frac{\avg{\mathbf{v}(t)\cdot\mathbf{v}(0)}}{\avg{v^2(0)}}
\end{equation}
where $\mathbf{v}(t)$ is the velocity of an atom at time $t$ and implicit in the ensemble average, $\avg{\cdots}$,
is an average over all atoms in the system.
The vibrational density of states is show in Fig.~\ref{fig:pwr} and displays two main features,
a high frequency peak and a low frequency peak.
The high frequency peak corresponds to the Cl-Cl stretch vibration.
All four functionals under study underestimate the frequency of the Cl-Cl stretch, which is experimentally
between 530~cm$^{-1}$ and 550~cm$^{-1}$ depending on the isotopic composition of the Cl$_2$ molecule~\cite{ClRaman}.
This underestimation of the stretching frequency
is consistent with each of the functionals predicting a Cl-Cl bond length that is too large,
and the functional dependence of this peak position
follows the bond lengths and their variances predicted by each functional, see Fig.~\ref{fig:rdf}b.
The functionals display significant differences in the shape of the low frequency peak in $I(\omega)$;
PBE+D3 and BLYP+D3 yield the same qualitative shape, while the SCAN-based functionals yield significantly
more density at higher frequencies.
The motions probed in this low frequency region of $I(\omega)$ involve collective rearrangements of the molecules
in the liquid, which are dictated by breakage and reformation of XBs.
Thus, we ascribe these differences to the presence of stronger XBs in the SCAN-based functionals
than PBE+D3 and BLYP+D3. 
%

%
To summarize this section, we find that SCAN+rVV10 provides the best representation of the structure of $l$-Cl$_2$,
at the cost of slowed dynamics. 
These slow dynamics arise from halogen bonds that are too strong, possibly due to
self-interaction and/or density-driven errors~\cite{Kim_2018,MHG_XB,XB-flosic}.
In contrast, the most accurate dynamics are predicted by PBE+D3, at the cost of a poorer overall description of $g(r)$.
In the remainder of this work, we quantify the structure and dynamics of halogen bonding of Cl$_2$.
Because a major focus is on dynamic properties of XBs, data discussed throughout the remainder
of the work is obtained using the PBE+D3 functional, and we make comparisons
to the SCAN+rVV10 functional where appropriate.
%

\section{Electronic Structure-based Definition of a Halogen Bond}
Halogen bonding is the result of electrostatic attractions between regions of high and low electron density
involving at least one halogen atom.
In order to understand the origin of halogen bonds, we first examine the electronic structure
of crystalline diatomic chlorine, whose crystal structure is a result of halogen bonding
and packing of lone pairs~\cite{Nyburg_PRSLA_1979,Steven_MolPhys_1979,Prince_MolPhys_1982,Tsirelson:br0033,Remsing:MolPhys:2018}.

\begin{figure}[tb]
\begin{center}
\includegraphics[width=0.43\textwidth]{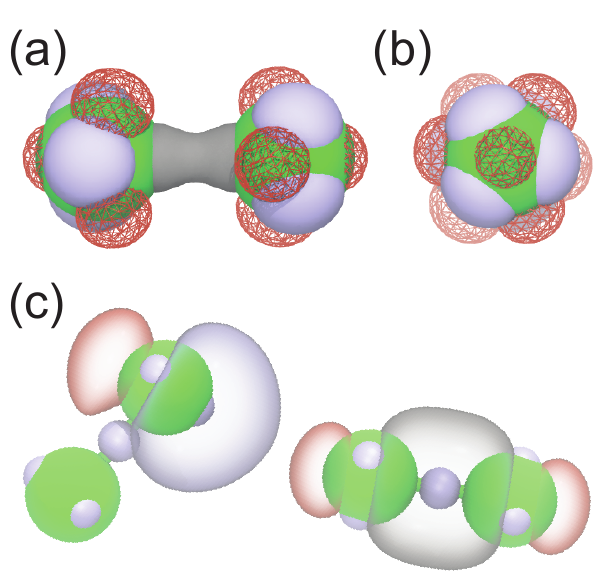}
\end{center}
\caption
{
(a,b) Maximally localized Wannier functions (MLWFs) of a Cl$_2$ dimer in the solid state,
shown from (a) the side and (b) down the Cl-Cl bond axis.
The solid gray isosurface indicates the Cl-Cl covalent bond and solid blue isosurfaces indicate lone pairs,
and both are drawn at a value of 0.27~Bohr$^{-3}$.
Red wireframe isosurfaces are opposite in sign to the solid surfaces and are drawn at a value of 0.09~Bohr$^{-3}$,
chosen to be three times smaller than the isodensity contour used for the solid surfaces for clarity.
Cl atoms are shown as green spheres.
(c) MLWFs involved in a halogen bond (XB) between two chlorine dimers in the solid state.
All isosurfaces are drawn at 0.05~Bohr$^{-3}$ following the same color scheme as in panels (a) and (b). 
Also shown are the centers of the MLWFs (MLWFCs) as small blue spheres.
Note that a XB between two dimers is consistent with a linear Cl-Cl-MLWFC arrangement.
}
\label{fig:mlwf}
\end{figure}

In Fig.~\ref{fig:mlwf}, we show the maximally localized Wannier functions (MLWFs) for a single Cl$_2$
in the solid, where blue and red isosurfaces indicate regions of high and low electron density, respectively,
and the covalent bond MLWF is highlighted in gray.
In dimeric chlorine, bromine, and iodine, electron density is depleted along the covalent bond axis, leading to the formation
of electron density deficient $\sigma$-holes at the ends of each dimer along the bond axis~\cite{Tsirelson:br0033,Remsing:MolPhys:2018,I2-2014}.
Similar $\sigma$-holes also develop between the lone pairs~\cite{Remsing:MolPhys:2018,I2-2014}.
These $\sigma$-holes can be readily observed as the wireframe regions of the MLWFs shown in Fig.~\ref{fig:mlwf}a,b.
A halogen bond forms when the lone pair region of one dimer
forms a Lewis-type interaction with the $\sigma$-hole of a neighboring dimer~\cite{I2-2014,Cavallo_2016}.
One such halogen bonding arrangement in crystalline Cl$_2$ is shown in Fig.~\ref{fig:mlwf}c,
along with the corresponding MLWFs involved in the XB and all the MLWF centers (MLWFCs)
of the two dimers.
The visualization in Fig.~\ref{fig:mlwf}c clearly indicates an electrostatic attraction between the
electron rich portion of the lone pair MLWF of the left molecule (blue surface)
with the electron deficient $\sigma$-hole of the right molecule (red surface). 
Moreover, Fig.~\ref{fig:mlwf}c suggests that halogen bonding is consistent with a linear
Cl-Cl-MLWFC arrangement.
Understanding the physical origin of XBs in this manner enables their quantification
through a MLWF-based approach.
In particular, we now introduce a geometric definition of a XB that is rooted in understanding the spatial correlations
among Cl atoms and MLWFCs.
%

\begin{figure}[tb]
\begin{center}
\includegraphics[width=0.43\textwidth]{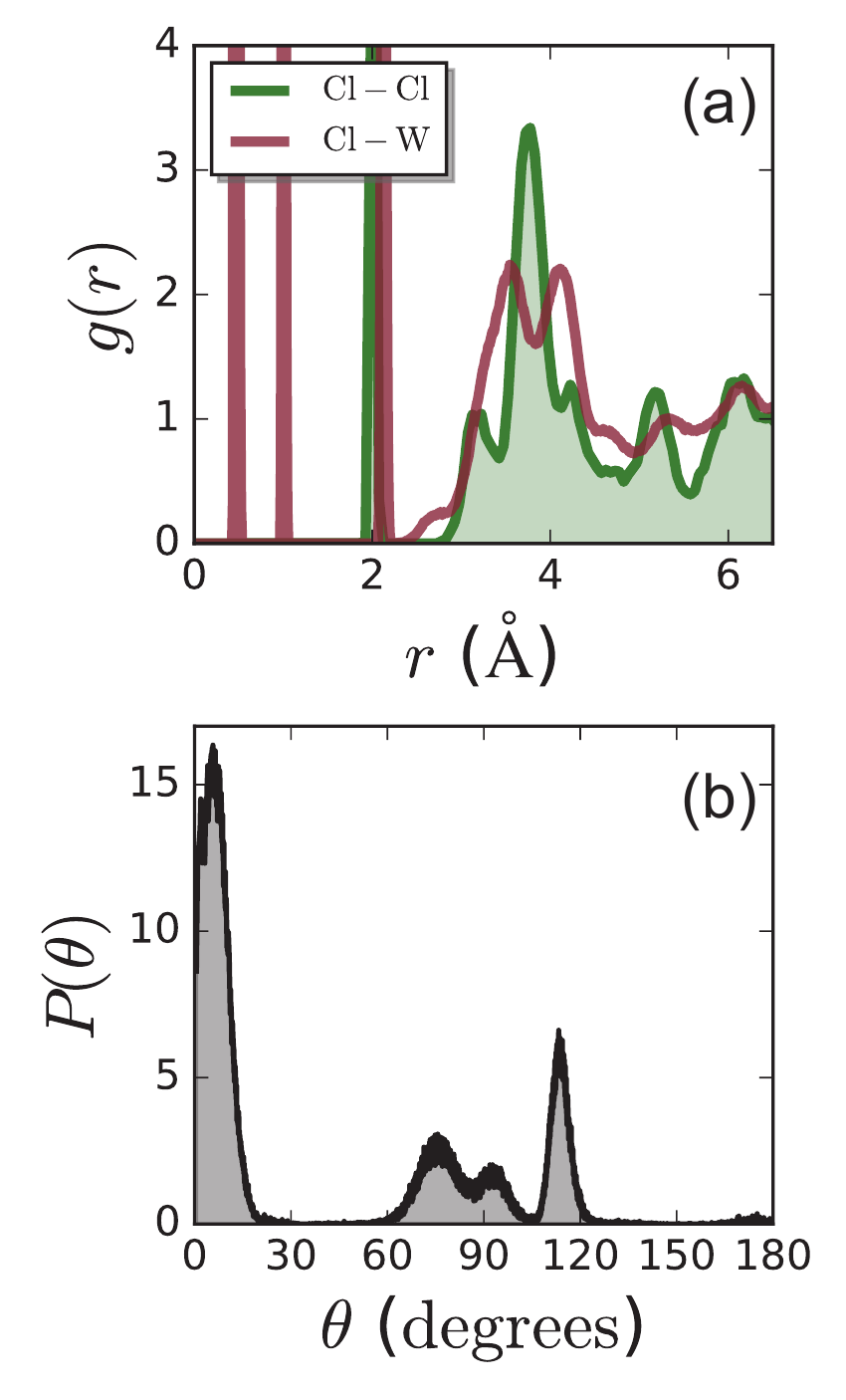}
\end{center}
\caption
{
(a) Radial distribution function, $g(r)$, for Cl-Cl and Cl-W (Cl-ML\textbf{W}FC) correlations in solid Cl$_2$ at
a temperature of 100~K.
(b) Probability distribution of the Cl-Cl-W angle for MLWFCs within a distance of 0.75~\AA \ of a Cl atom.
}
\label{fig:rdf}
\end{figure}

%
The first component of our halogen bonding criterion is a Cl-Cl distance cutoff that defines a maximum
distance for which two Cl atoms can be considered halogen bonded.
The Cl-Cl radial distribution function, $g(r)$, shows a sharp peak at $r\approx2$~\AA \ that corresponds
to the covalent bond in molecular chlorine, Fig.~\ref{fig:rdf}a.
The $g(r)$ then displays several peaks between $r\approx3$~\AA \ and $r\approx4.5$~\AA,
the first of which is indicative of halogen-bonded Cl-Cl contacts.
Thus, we define our distance cutoff based on the first minimum following this peak,
such that $r_{\rm ClCl}<3.4$~\AA.
%

\begin{figure}[tb]
\begin{center}
\includegraphics[width=0.43\textwidth]{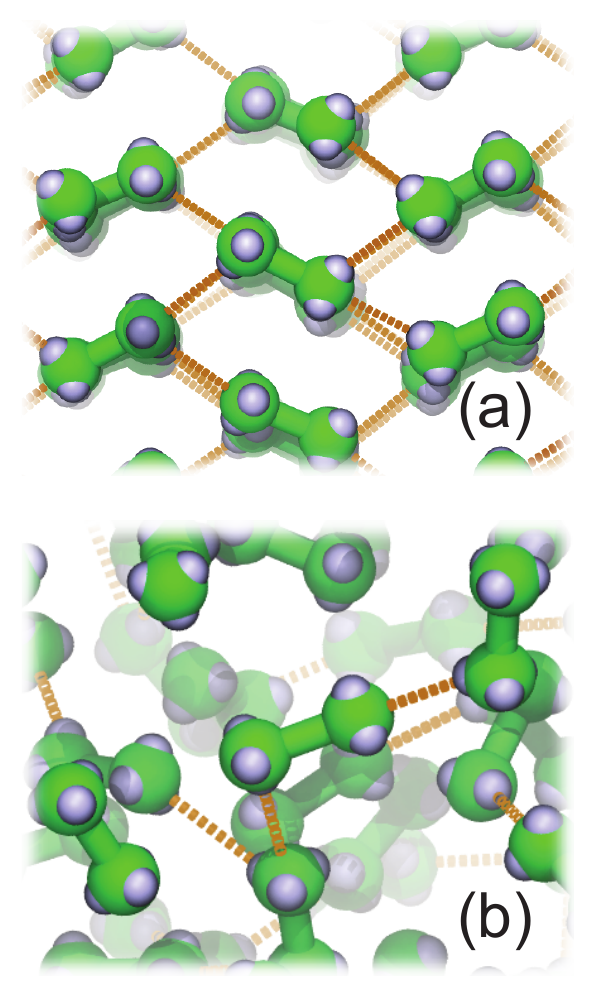}
\end{center}
\caption
{
Snapshots illustrating halogen bonds in (a) solid and (b) liquid Cl$_2$, at 100~K and 200~K, respectively.
Cl atoms are the large green spheres, MLWFCs are the small blue spheres,
and halogen bonds are indicated by the orange dashed cylinders and identified
according to the geometric criterion described in the text.
}
\label{fig:snap}
\end{figure}

%
We now include a second component of the geometric criterion for XBs that includes
correlations involving MLWFCs.
The $g(r)$ characterizing correlations between MLWFCs and Cl atoms (Cl-W, where W indicates a ML\textbf{W}FC)
is also shown in Fig.~\ref{fig:rdf}a.
There are two types of MLWFCs in Cl$_2$, lone pair and covalent bond MLWFCs.
The lone pair MLWFCs correspond to the first peak in the Cl-MLWFC $g(r)$ near $r_{\rm ClW}\approx0.5$~\AA,
as well as the sharp peak near $2.1$~\AA.
The covalent bond MLWFC contributes to the peak near $1$~\AA, roughly half the Cl-Cl bond length.
The peak just before $3$~\AA \ is also consistent with lone pair MLWFCs between
two Cl atoms in a linear halogen bonding configuration.
A halogen bond is defined by a linear Cl-W$\cdots$Cl arrangement, where the MLWFC (W) here
corresponds to a lone pair.
The probability distribution, $P(\theta)$, of the Cl-Cl-W angle, for MLWFCs within a distance of 0.75~\AA \ of a Cl atom,
is shown in Fig.~\ref{fig:rdf}b.
The distribution $P(\theta)$ shows a large peak near $\theta=0\degree$, indicative of XBs.
Additionally, there are peaks near $75\degree$, $92\degree$, and $115\degree$,
corresponding to MLWFCs that are not involved in a
XB with either of the Cl atoms in the Cl-Cl-W triplet.
The sharp non-XB peak near $\theta=115\degree$ corresponds to MLWFCs that are on the same Cl atom as
the MLWFC involved in an XB.
The remaining peaks corresponds to MLWFCs on the other Cl atom in the triplet, which
is participating in the XB via a $\sigma$-hole.
We summarize our halogen bonding criterion as follows.
A halogen bond between two Cl atoms exists if $r_{\rm ClCl}<3.4$~\AA \ and
the Cl-Cl-W angle is $\theta<30\degree$, such that the MLWFC in the triplet corresponds to
a lone pair, \emph{e.g.} it is within 0.75~\AA \ of one of the Cl nuclei.
Examples of XBs in solid and liquid Cl$_2$ using our criterion are shown in Fig.~\ref{fig:snap}.
%

%
We conclude this section with a discussion of the robustness of our approach with respect to
traditional XB definitions.
Halogen bonds are often defined using nuclear coordinates only.
In the case of a XB between two Cl$_2$ molecules,
the halogen bond would be defined using the Cl-Cl intermolecular distance
and the two angles formed by the Cl bond vectors and Cl-Cl intermolecular distance vector.
While useful, such definitions include no information about the electronic structure of the system.
By including information about the electronic degrees of freedom, our proposed
definition is able to accurately and robustly characterize XBs in molecular systems,
including situations where the purely nuclei-based definitions fail.
%

\section{Halogen Bonds in Solid and Liquid Chlorine}
\begin{figure}[tb]
\begin{center}
\includegraphics[width=0.48\textwidth]{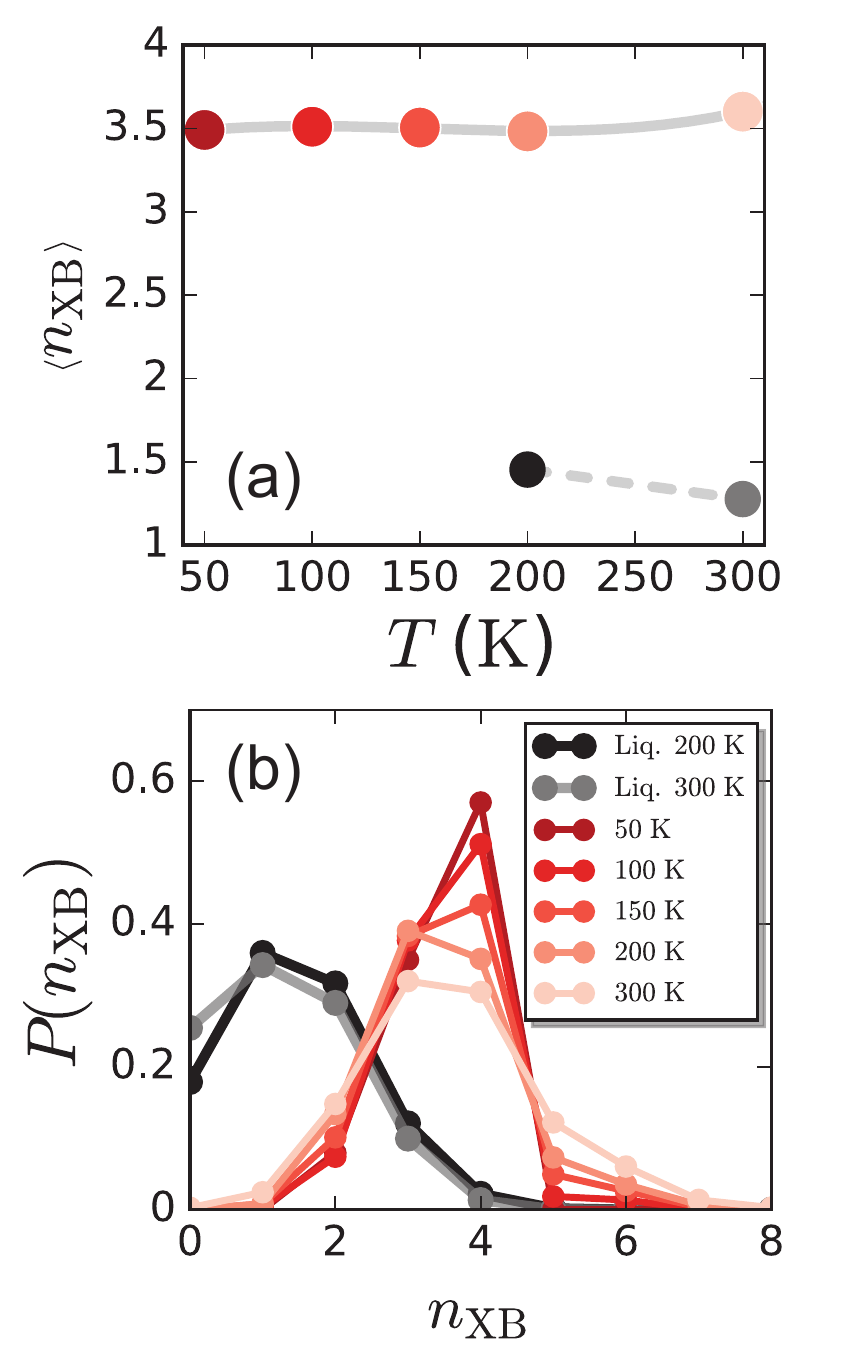}
\end{center}
\caption
{
(a) Average number of halogen bonds per molecule, $\avg{n_{\rm XB}}$, along an isochore in solid Cl$_2$ (points on solid line)
and in liquid Cl$_2$ (points along dashed line). 
Lines are guides to the eye. 
(b) Probability distribution of the number of halogen bonds per molecule, $P(n_{\rm XB})$,
for the state points in panel a.
}
\label{fig:nhb}
\end{figure}

We can use the XB definition in the previous section to characterize the statistics
of XBs in solid and liquid Cl$_2$.
The average number of XBs per molecule, $\avg{n_{\rm XB}}$,
is approximately 3.5 in the solid state; see Fig.~\ref{fig:nhb}a.
In the solid, each Cl atom can donate and accept a XB, as shown in Fig.~\ref{fig:snap}a,
with thermal fluctuations transiently disrupting these interactions and reducing $\avg{n_{\rm XB}}$ to 3.5,
from the ideal value of 4.
The average number of XBs per molecule
reduces to approximately 1.5 upon melting at 200~K, and further reducing to
1.3 at 300~K, as shown in Fig.~\ref{fig:nhb}a.
This reduction in halogen bonding is consistent with the lower density of the
liquid --- 12.5~molecules/nm$^3$ at 200~K and 8.32~molecules/nm$^3$ at 300~K
as compared to approximately 18~molecules/nm$^3$ in the solid ---
as well as the increased rotational and translational dynamics of chlorine molecules.
%

%
We also examine the probability distribution of the number of XBs per molecule, $P(n_{\rm XB})$, Fig.~\ref{fig:nhb}b.
In the solid state, $P(n_{\rm XB})$ is peaked around $n_{\rm XB}=4$, consistent with the expectation that
each Cl atom can donate and accept a XB in the orthorhombic arrangement of the solid.
These XBs generally
lead to the unique crystal structure of the larger halogen dimers, Cl$_2$, Br$_2$, and I$_2$~\cite{Price_Chlorobenzene,Nyburg_1968,Prince_MolPhys_1982,Tsirelson:br0033,I2-2014,Remsing:MolPhys:2018}.
As the temperature is increased along an isochore,
the width of $P(n_{\rm XB})$ increases, due to increased fluctuations of the crystal lattice,
and the maximum shifts to $n_{\rm XB}=3$ in the superheated states, $T>171$~K.
In the liquid state, $P(n_{\rm XB})$ is peaked at $n_{\rm XB}=1$, and exhibits significant probably at values of
$n_{\rm XB}$ between 0 and 4 XBs per molecule.
The broad distribution of XBs in the liquid suggests that there is significant structural heterogeneity
in $l$-Cl$_2$, which is not present in the solid state.
This structural heterogeneity leads to broader distributions of observables, such as the larger linewidth
of the Cl-Cl stretching vibration peak in the Raman spectra of $l$-Cl$_2$, as compared to that of the solid.
\\

\section{Halogen Bond Dynamics in Chlorine}
The MLWFC-based XB definition used here enables the characterization
of XB dynamics.
In particular, we define an indicator function, $h(t)$, which is equal to one when a XB
exists at time $t$ between two atoms and zero otherwise.
Halogen bond dynamics can then be probed with the time correlation function (TCF)
\begin{equation}
C(t)=\frac{\avg{h(t)h(0)}}{\avg{h}},
\end{equation}
in analogy with the procedure often used to probe hydrogen bonding dynamics~\cite{LuzarChandler,Luzar_2000,Kumar_2007,Laage_Sciene_2006,Laage_2008}.
%

\begin{figure}[tbh]
\begin{center}
\includegraphics[width=0.45\textwidth]{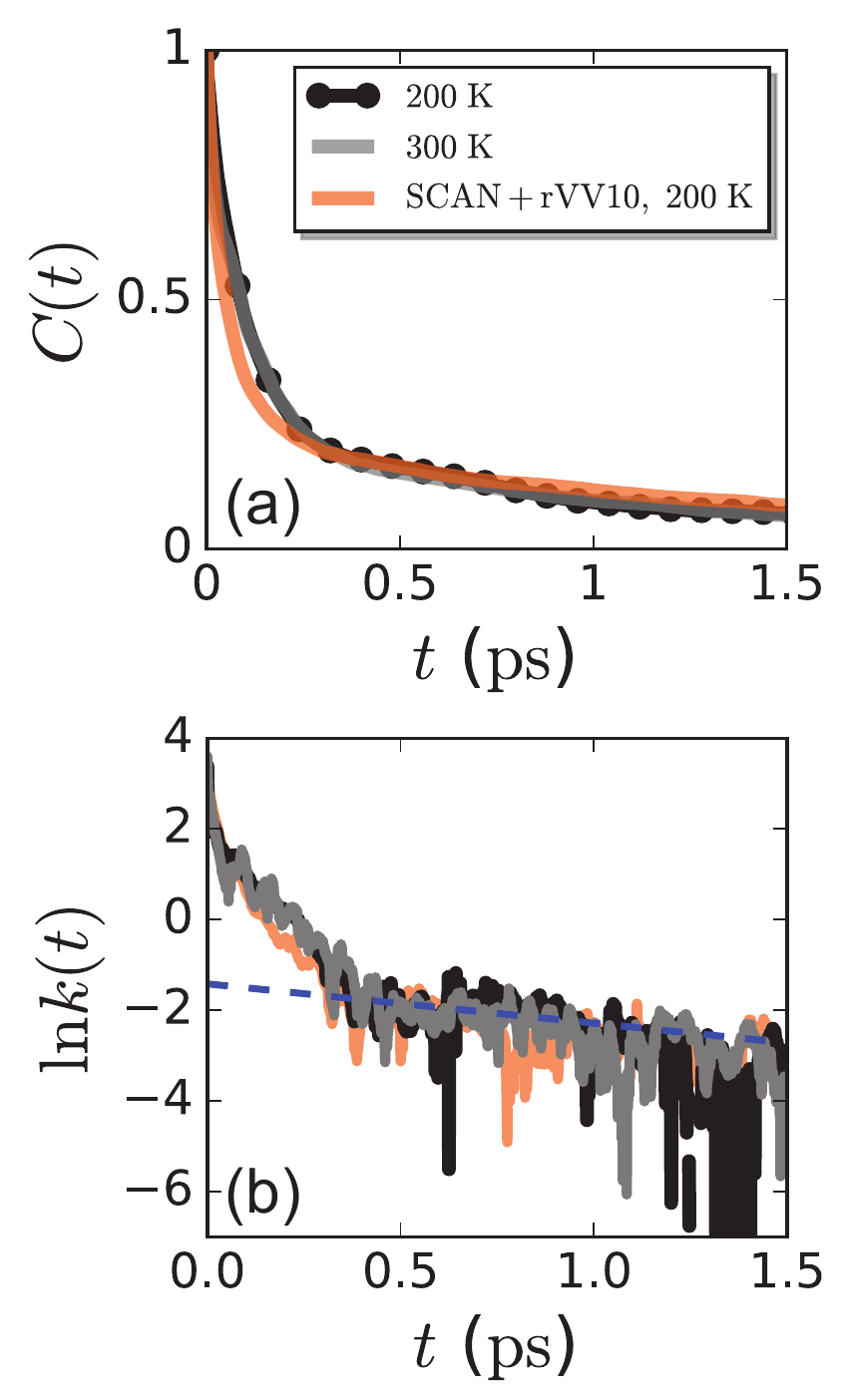}
\end{center}
\caption
{
(a) Halogen bond time correlation function, $C(t)$, for liquid Cl$_2$ at 200~K (black) and 300~K (gray),
as well as that predicted by SCAN+rVV10 at 200~K (orange).
(b) Corresponding reactive flux correlation functions, $k(t)$, and a fit
to $k(t)\sim\tau^{-1} \exp(-t/\tau)$ at long times (dashed lines), where $\tau$ is the halogen bond lifetime,
is shown for the 200~K results.
}
\label{fig:TCF}
\end{figure}

%
The XB TCF is shown in Fig.~\ref{fig:TCF}a for $l$-Cl$_2$ at 200~K and 300~K.
The decay of $C(t)$ can be fit by a biexponential decay with time scales
$\tau_1\approx0.08$~ps and $\tau_2\approx1.3$~ps at $T=200$~K.
We also compute the reactive flux correlation function~\cite{Chandler_1978,LuzarChandler,Luzar_2000},
\begin{equation}
k(t)=-\frac{d C(t)}{dt}=-\frac{\avg{\dot{h}(0)\brac{1-h(t)}}}{\avg{h}},
\end{equation}
which plateaus to a value of $k(t)\sim \tau^{-1} \exp(-t/\tau)$ after an initial transient period, as shown in Fig.~\ref{fig:TCF}b.
Indeed, fitting of $k(t)$ to the expected form in the plateau region yields a halogen bonding timescale of $\tau\approx1.15$~ps
at 200~K, in agreement with the biexponential decay of $C(t)$.
The SCAN+rVV10 functional yields a longer XB lifetime, $\tau\approx1.23$~ps at 200~K,
consistent with the stronger halogen bonds in this system, although the initial transient decay is faster
than that predicted by PBE+D3.
At 300~K, the XB lifetime shortens to $\tau\approx1.07$~ps, as may be expected from the increased
dynamics at higher temperatures.
The change in $\tau$ from 200~K to 300~K closely tracks the change in the rotational relaxation time,
$\tau_2\approx1.13$~ps at 200~K and $\tau_2\approx1.07$~ps at 300~K, as determined
by fitting the long-time behavior of $C_2(t)$ to an exponential decay.
This correlation between $\tau$ and $\tau_2$ supports the earlier suggestion
that a significant pathway for rotational relaxation in
$l$-Cl$_2$ involves XB breakage.
We additionally note that the XB lifetime in $l$-Cl$_2$ in this temperature range is on the order of a picosecond, 
similar to the lifetime of hydrogen bonds in water at ambient conditions~\cite{LuzarChandler,Luzar_2000}.
The computation of $C(t)$ shown here demonstrates that dynamic properties of XBs
can be readily evaluated using our approach.
For example, the time-dependence of XBs in contexts such as halogenated ligand unbinding
from proteins and phase transitions in supramolecular assemblies can be readily quantified
and halogen bonding rate constants can be computed.
These concepts will shed light on the role of XBs in
determining the kinetics of a wide array of processes in the chemical, materials, and biological sciences.
%

\section{Conclusions}
In this work, we have used ab initio molecular dynamics simulations in combination
with the maximally localized Wannier function formalism to characterize the structure and dynamics
of condensed phase halogen bonds on a footing equal to traditional measures of hydrogen bonding.
This consistent picture of noncovalent, directional interactions enables extension of the vast literature on
hydrogen bonding in molecular systems to characterize halogen bonding.
We close with a discussion of how our results may be used to develop classical,
empirical models of halogen bonding, which will enable molecular simulations on larger length and time scales.
Such models will be important for describing
halogen bonding in supramolecular assemblies
and protein-ligand complexes, for example,
especially if dynamic and thermodynamic properties are of interest.
We expect that an empirical model of halogen bonding in Cl$_2$ can
be developed from first principles using the insights provided by our XB analysis scheme.
In particular, one might imagine constructing a semi-rigid, 8-site model of Cl$_2$, wherein each Cl atom
is represented by four sites, one Cl nucleus and three lone pair sites, (LP).
This differs from recently developed empirical models of halogen bonding in that the lone pair
sites are explicitly represented~\cite{Jorgensen_Review,Jorgensen_JCTC,Kol__2016}.
Bond lengths and angles involving Cl and LP sites could be determined from AIMD averages,
and
the charges on the LP and Cl sites may be chosen to reproduce the quadrupole moment of the Cl$_2$ molecule,
or tuned to match the structure of condensed phase Cl$_2$ more accurately.
Alternately, the intermolecular interactions could be developed through machine learning approaches applied
to ab initio computations of the type reported herein~\cite{PhysRevMaterials.3.023804,PhysRevLett.120.143001}.
Moreover, we expect such empirical representations of halogen bonding to be transferable
to halogenated compounds in general, including organic crystals and biomolecular systems.
%

\begin{acknowledgements}
This work was supported as part of the
Center for Complex Materials from First Principles (CCM), an Energy Frontier
Research Center funded by the U.S. Department of Energy, Office of Science, Basic Energy
Sciences under Award \#DE-SC0012575.
Computational resources were supported in part by the National Science Foundation
through major research instrumentation grant number 1625061
and by the US Army Research Laboratory under contract number W911NF-16-2-0189.
\end{acknowledgements}

\end{document}